\documentstyle[preprint,aps]{revtex}
\begin{document}
\begin{titlepage}
\begin{center}
\today     \hfill    MIT-CTP-2464 \\
           \hfill    hep-ph/9509212 \\
\vskip .3in
{\large {\bf Challenging weak-scale supersymmetry at colliders}}
\vskip .3in
Greg W. Anderson\footnote{Current address: 
Theory Department, MS106, Fermi National Accelerator Laboratory,
PO Box 500, Batavia, IL 60510 USA.}
and Diego J. Casta\~no 
~\footnote{Current Address: Dept. of Physics, Florida State University,
Tallahassee, FL 32306 USA.}\\
{\em Center for Theoretical Physics, Laboratory for Nuclear Science \\
 Massachusetts Institute of Technology\\
Cambridge, MA 02139}\\
\end{center}
\begin{center}
To appear in: {\it Phys. Rev. D}
\end{center}
\vskip .20in
\begin{abstract}
Experimental searches for supersymmetry are entering a new era.
As future experiments explore the mass range above the current 
lower bounds on superpartner masses, 
a failure to observe 
signals of superpartner production will begin
to erode the central motivation for supersymmetry at the weak
scale.  In this article we present a detailed examination of 
which regions of supersymmetric parameter space are most natural and
the extent to which weak-scale supersymmetry becomes unnatural
if no superpartners are observed at LEP-II, the Tevatron, possible
upgrades of these machines, and the LHC.

\end{abstract}

\end{titlepage}
\renewcommand{\thepage}{\roman{page}}
\setcounter{page}{2}
\mbox{}
\renewcommand{\thepage}{\arabic{page}}
\setcounter{page}{1}

\section{Introduction}
\def\theequation{1.\arabic{equation}}
\setcounter{equation}{0}
\indent

Supersymmetry's continued growth in popularity as a candidate for 
physics beyond the standard model can be traced to two sources.
First, there is suggestive, circumstantial evidence for weak-scale
supersymmetry.  The lightest superpartner (LSP) makes an attractive
dark matter candidate, and the measured value of $\sin^2\theta$ 
coincides with the prediction of supersymmetric grand unification.
A second, and no less influential, factor is the
relative ease with which supersymmetry maintains consistency, both with
precision tests of the standard model and with the failure to observe
new particles in collider experiments, when compared to many of 
its competitors.  
This is because supersymmetric deviations from the standard model 
 can be made arbitrarily small by simply raising the masses of 
superpartners.

However, the search for weak-scale supersymmetry has 
reached a milestone.   Although previous
experiments could raise the lower bound on superpartner masses
without posing a serious challenge to supersymmetry, 
further increases in this bound 
will begin to erode a dominant motivation for supersymmetry at 
the weak scale.  
The central motivation for the appearance of supersymmetry 
at the weak scale is its ability to solve the naturalness
problem \cite{Witten}.  
In the standard model, the mass of the Higgs boson is
quadratically divergent.  If the standard model gives a complete
description of nature below a scale $\Lambda$,  quadratically 
divergent contributions to the square of the Higgs boson mass must 
cancel against bare terms to a part in $m_{H}^2/\Lambda^2$.  
If $\Lambda$ is large, 
precise cancellations must be engineered at every order 
in perturbation theory.  Without additional structure or 
organizational principles beyond the standard model, it would be
unnatural if the Higgs mass were much lighter than the cutoff,
since the cancellations that ensure a light Higgs mass would be 
upset by a variety of minute changes in the fundamental parameters.

Supersymmetry allows scalar masses to remain light in comparison to 
a cutoff without delicate fine-tuning because the 
additional  
renormalization effects of superpartners cancel the quadratic 
divergences in a supersymmetric theory.  
However, because supersymmetry is broken, these cancellations are
only achieved up to the mass splittings between the standard model particles
and their superpartners.  In this way, the scale of supersymmetry
breaking assumes the role of the cutoff in the previous discussion.
Accordingly, as superpartner masses
increase, a point is reached where supersymmetry is no longer
able to provide a complete explanation of why a light weak
scale is natural.  

Recently, we constructed a family of naturalness measures which 
reliably quantify how unnatural the supersymmetric standard 
model becomes as the masses of superpartners increase
\cite{Fine_Tuning}.  
These naturalness measures provide a significant advance 
over popular characterizations of naturalness.  Previous attempts
to quantify naturalness \cite{Sensitivity}  
are known to overestimate fine-tuning by
an order of magnitude or more \cite{Fine_Tuning}.  Other less quantitative
criteria are also problematic.
For example,
it is often loosely stated that superpartners must lie 
``at or below 1 TeV''
if the weak scale is to arise naturally, or that superpartner masses
should not lie ``too far above the weak scale.''  
As defining rules for what constitutes naturalness,
these compact statements are less than adequate.
First, there are dramatic differences between
the upper bounds on the masses of different superpartners.  If we are 
willing to tolerate solutions with $10$ $(20)$ times more fine-tuning 
than ideally natural solutions require,
squark masses may be as heavy as $750$ $(1200)$ GeV.  
By contrast, under this same
criterion, the mass of the lightest 
neutralino cannot exceed $145$ $(250)$ GeV.
Second, the often quoted range of superpartner masses
``at or below $1$ TeV'' ignores the 
{\em progressive}  worsening of the naturalness
problem as superpartner masses are increased from the weak scale
to 1 TeV. 
A scenario with any superpartner mass as heavy 
$1$ TeV is significantly less natural than a variety of lower 
mass solutions.  For example, a $1$ TeV squark mass requires adjustments
in the fundamental parameters with 16 times more 
precision than a 200 GeV squark mass.  Moreover, for some types 
of superpartners, a mass of $1$ TeV  would be extremely unnatural.  The
mass of the lightest chargino can only reach $1$ TeV 
at the expense of a factor of $45$ in fine-tuning, 
and an LSP of mass $1$ TeV requires fine-tuning 
to at least a part in $65$.

  In this paper we attempt to provide a detailed picture
of what regions of supersymmetric, parameter space are most natural, 
where we should have the strongest expectations of observing
signals of supersymmetry, and how 
supersymmetry's ability to accommodate a light weak scale
naturally will be progressively challenged
if signals of superpartner production are not observed in the future.
After reviewing our quantitative naturalness measures, we 
construct naturalness contours and compare them
to the SUSY discovery reach of current and future collider experiments.

\section{Quantifying Naturalness }
\def\theequation{2.\arabic{equation}}
\setcounter{equation}{0}
\indent

  This section describes the quantitative methods we use to determine
naturalness in the minimal supersymmetric model (MSSM).  A detailed
derivation of these naturalness measures and example applications
can be found in Ref. [2].     
If we parametrize our assumptions about
the likelihood distribution of a theory's fundamental parameters,
a fine-tuning measure can be constructed directly from probability 
arguments, and we can retain a functional
parametrization of our assumptions that can be used to 
estimate the theoretical uncertainty in the naturalness measure.
Consider an effective field theory with a set of measurable
parameters $X$.  Using the renormalization group, 
the $X$'s can be written in terms of 
the presumably more fundamental, high-energy boundary conditions of 
the effective theory.
If we write the probability distribution for these `fundamental 
parameters' as
\begin{equation}
 dP = f(a)da,
\end{equation}
a naturalness measure $\gamma$ can be  
constructed from the following prescription:  Write the observables
in terms of the fundamental parameters $X=X(a)$,
compute the probability that $X$ lies within a specified interval
about $X$,  average this probability over the likelihood
distribution of the fundamental parameters,
and then divide this average by the probability that
$X$ lies within the specified interval about a particular 
value of $X$ \cite{Fine_Tuning}. With this prescription, ideally natural solutions 
will have $\gamma = 1$, while finely tuned solutions yield 
$\gamma >> 1$.
If we define a
sensitivity function $c(X,a) = |(a/X)(\partial X /\partial a)|$,
and if we define an average sensitivity $\bar{c}$ by
\begin{equation}
1 /\bar{c}  = \frac{\int da \, a f(a) \, c(X;a)^{-1} }
{a f(a)\int da},
\end{equation}
the naturalness measure can be written
\begin{equation}
\gamma = c/\bar{c} \ .
\end{equation}
As suggested by Eq. (2.3),
the naturalness measure $\gamma$ 
is equivalent to a refined version of Wilson's naturalness 
criteria \cite{Wilson}: 
Observable properties of a system,
{\it i.e.,} $X$, should not be {\it unusually unstable} with respect
to minute variations in the fundamental parameters, $a$.  

 
  We will use this measure to study the naturalness of the weak
scale as the masses of superpartners are increased.
In the MSSM, electroweak symmetry breaking is induced
by radiative effects.  The Z-boson mass serves as a useful order
parameter for electroweak symmetry breaking because it has
a relatively simple dependence on the Lagrangian parameters:
\begin{equation}
{1 \over 2}m^{2}_{Z} = 
{ {\overline m}^{2}_{\Phi_d} -{\overline m}^{2}_{\Phi_u} \tan^2\beta 
\over \tan^2 \beta -1} - \mu^2.
\end{equation}  
The parameters ${\overline m}^2_{\Phi_d}$ and ${\overline m}^2_{\Phi_u}$ 
are the quadratic mass terms for the two Higgs doublets with 
additional terms from the
one-loop corrections to the effective potential \cite{Fine_Tuning}.
The $\mu$ parameter is a coupling between the two different Higgs doublets,
and $\tan\beta$ is the ratio of their vacuum expectation values.  

In our study we focus on supersymmetric extensions of the standard
model compatible with minimal supergravity and unification
for which the soft supersymmetry breaking can be written
in terms of four parameters $A, B, m_0$, and $ m_{1/2}$.  
At low energies, the values of
${\overline m}^{2}_{\Phi_d}$,${\overline m}^{2}_{\Phi_u}$, 
$\mu$, and $\tan\beta $ can be written in terms of 
these four soft SUSY-breaking parameters and
other couplings at the high-energy scale such as the top 
quark Yukawa coupling $y_t$.  
We can evaluate the naturalness of the weak scale
with respect to any of these parameters.
To accurately asses the naturalness of a particular solution
it is necessary to compute the fine-tuning
with respect to several parameters.~\footnote{Computing the fine-tuning
with respect to only one of the fundamental parameters
leads to a significant underestimate of fine-tuning.}
Unnatural solutions 
are typically only fine-tuned with respect to a few of the fundamental
parameters, and which parameters are fine-tuned will vary from solution
to solution.  For our purposes it is sufficient to consider the naturalness
with respect to two dimensionful parameters and one dimensionless coupling.
For each solution we define a cumulative naturalness measure by
\begin{equation}
\tilde{\gamma} = {\rm max}
\left\{ \gamma(y_t),\gamma(m_0),\gamma(m_{1/2})  \right\}.
\end{equation}
To evaluate $\gamma$ we must assume a distribution for the
fundamental parameters $f(a)$ and a suitable range of 
integration.  We will make two choices for the distribution
$f_{1}(a)=1$ and $f_{2}(a) = 1/a$.  The difference between
these respective $\gamma$'s is a measure of the theoretical uncertainty
in naturalness contours.  Inspection of Figs. 1a-b shows that this
theoretical uncertainty is much less than order one, and typically lies
in the range of a few to tens of percents.

For each distribution, we average the sensitivity of the weak scale over 
a range of the fundamental parameters $a$ following the method of 
Ref. \cite{Fine_Tuning}.  
Since we employ two-loop 
renormalization group equations (RGE's) for all parameters (except the 
soft breaking ones), numerical methods are used to solve the RGE's 
and to integrate (2.2).  The procedure is time consuming, especially 
for cases in which the range of $a$ is large.  
We evaluated the naturalness of approximately $4,000$ different 
solution sets from which we constructed the naturalness contours
in Figs. 1-5.  To efficiently determine these contours, we selectively 
explored the more pertinent regions of parameter space.  
Because the data points used to construct our naturalness contours
were scattered over a large range of the fundamental parameters,
no particular choice of $A$, or $\tan\beta$ is implicit in the 
naturalness contours found in Figs 1-5.  Particular choices for 
$A$ or $\tan\beta$ would result in curves that would be more 
restrictive. 
Figs. 2-5 compare the more conservative naturalness measure
$\tilde{\gamma}_2$, to various estimates of SUSY discovery reaches.
Each comparison is made both in terms of the gluino and squark masses
(Figs. 2a-5a),
and in terms of the soft SUSY breaking parameters $m_{1/2}$ and
$m_0$ (Figs. 2b-5b).


\section{Challenges to SUSY at Colliders }
\def\theequation{3.\arabic{equation}}
\setcounter{equation}{0}

  At particle colliders, 
sparticles must be produced in pairs to conserve R-parity. 
Once produced, a superpartner decays through a cascade ending in a stable, 
unobservable LSP.  Accordingly, signals of superpartner production at
colliders  would consist of events with missing  energy 
and various combinations of leptons and jets. 
The CERN $e^{+}e^{-}$ collider LEP will soon be upgraded in
energy to begin phase II.  LEP-II is expected
to start a trial run just below the $WW$ threshold in late 1995,
and by 1996 it is expected to achieve energies of
$\sqrt{s}=(175-190)$ GeV .  
Potential searches for supersymmetry at LEP-II have 
been studied by many groups \cite{LEPIIa,LEPIIb,LEPIIc}.  Typically,
chargino pair-production and slepton pair-production are the two
most promising channels for discovering supersymmetry at LEP-II,
however in some cases neutralino pair-production can provide the largest
reach.
The heaviest reach in slepton masses will come from selectron searches
since, unlike staus and smuons, selectron pairs can be produced by 
t-channel neutralino exchange in addition to the s-channel production 
mediated through a $\gamma$ or Z.   For small $m_0$, 
selectron searches should probe up to nearly the kinematic limit,
reaching masses of 82 (88) ((96)) GeV for center of mass energies of
175 (190) ((205)) GeV \cite{LEPIIc}.
For large $m_0$, searches for charginos should probe
masses up to 87 (95) ((102)) GeV for center of mass energies of
175 (190) ((205)) GeV.

In Figs. 2-3 we compare the SUSY discovery
reach in the selectron and chargino channels at LEP-II with
Tevatron search reaches and naturalness contours.
\footnote{We caution the reader that estimates
of search reaches should be interpreted carefully.  
A global determination of discovery reaches is a complicated 
undertaking.  The Tevatron search reaches quoted here have 
only been calculated for a few selected points in SUSY parameter 
space (for recent work on a more global analysis of search reaches
see Ref [12]).
Moreover, there is
sometimes significant differences between the reach estimates of
different groups, and the various background cuts in these
analyses have not been optimized to yield the maximum
reaches.}
The dashed curves $\tilde{e}(88)$ and $\tilde{\chi}^{+}(95)$
represent the approximate selectron and chargino mass reaches 
achievable with $\sqrt{s}=190$ GeV at LEP-II.  
Because the reaches in these channels depend on 
$\tan\beta$ and the sign of $\mu m_{1/2}$, 
we have chosen values of $\tan\beta$ which represent
larger and smaller reaches for each case.  Our convention
for the sign of $\mu$ agrees with Ref. [9,10].

Fermilab's $p\bar{p}$ collider, the Tevatron, will complete
the second stage of run I at the end of this year.
Through the end of run I, the classic missing transverse energy
and jet events (\hbox{/\kern-.6500em $E_{T}$})
 will provide the largest reach into SUSY parameter
space.
The dotted lines in Figs. 2-5 display the mass
limits on squarks and gluinos 
achieved by the Collider Detector at Fermilab (CDF) and D0 
experiments during run IA \cite{CDFD0}.
By the end of the current Tevatron run, run IB, the CDF and D0 
experiments can each be expected to have over 100 $pb^{-1}$ 
of data on tape.
Run II, with the Tevatron's Main injector upgrade (MI), is anticipated 
for 1999.  The MI run is expected to achieve ($\sqrt{s} =2$ TeV) and a 
yearly integrated luminosity of $\sim 1\, fb^{-1}$ per experiment.
Figs. 2a-b, compare naturalness contours to the discovery reach of 
missing energy and jet events during runs I and II,  as estimated by 
three different analyses \cite{BKT,BCKT,KLMW,DPF}.  
Although there is not always a 
clear consensus among these search reach estimates, 
several features are apparent.  First, if the weak scale is to 
arise naturally, 
the fundamental soft SUSY breaking parameters $m_0$ and $m_{1/2}$
should be significantly lighter than 1 TeV.
Accordingly, it is very unlikely that $m_{\tilde{q}}>>m_{\tilde{g}}$ 
unless the gluino is extremely light ($<1$GeV)
and has somehow managed to escape detection \cite{GF}.
Second, to set the most significant bounds from missing energy and jet 
events, the CDF and D0 experiments should pay special attention to
the regime $m_{\tilde{q}}\simeq m_{\tilde{g}}$.  
In addition, searches for chargino and selectron production at LEP-II
will cover a significant amount of the natural region 
of parameter space, particularly for $\mu > 0$.

Although missing energy and jet events will provide the Tevatron's 
strongest probe of supersymmetry during run I, searches in
other channels will become important in the main injector era.
During Tevatron run II,
missing energy events with same sign di-leptons 
and tri-leptons (3l) will provide a larger reach into parameter space
\footnote{ We caution the reader,
that these signals, particularly the tri-lepton events, 
have a strong dependence on $\tan\beta$ and the sign of $\mu$.
For this reason the discovery reaches in Figs. 3 and 5
can not be taken as global reaches.} 
than the background limited \hbox{/\kern-.6500em $E_{T}$}
channel for
some values of $\tan\beta$ and $\mu$ \cite{BKT,BCKT}.
Figs. 3 and 5 compare the physics
reach of the tri-lepton events, for $\mu <0$ and $\tan\beta=2$,
to the discovery reach of LEP-II and naturalness contours.
While LEP-II provides its largest reach into SUSY parameter 
space for $\mu >0$, Tevatron searches in the $3l$ channel
are most effective if $\mu <0$.  Moreover,
for $\mu<0$, the chances for observing 
tri-lepton signals at the Tevatron are promising
for run II.  Except for a small window where the branching ratio
of the second heaviest neutralino into
an invisible neutrino 
($\tilde{\chi}^{0}_{2}\rightarrow\nu \tilde{\nu}$)
approaches  $100\%$,  the reach in the tri-lepton
channel is largest for the most natural region of parameter space.

Further luminosity upgrades of the Tevatron, dubbed the TeV*, and
the Di-Tevatron have  been considered.
The TeV* upgrade, achievable through anti-proton
recycling and storage, could reach integrated luminosities of 
$25 pb^{-1}$, and the use of SSC type magnets could allow a
Di-Tevatron upgrade to achieve $\sqrt{s}=4$ TeV.
In Figs. 4-5, we compare the SUSY discovery prospects 
of these proposed
Tevatron upgrades with LEP-II operating at $\sqrt{s}=205$ GeV.
The dashed curves in Figs. 4-5 labeled
$\tilde{e}(96)$ and $\tilde{\chi}^{+}(102)$, represent the
approximate mass reach of LEP-II in these channels \cite{LEPIIc}.
From Figs. 4a-b we see that searches in the 
\hbox{/\kern-.6500em $E_{T}$} channel
at a Di-Tevatron would cover most of the
natural region of SUSY parameter space, 
while searches in the $3l$ channel would be very promising
at the TeV* if $\mu <0$.

  Finally we mention that the non-observation of a significant
amount of missing transverse energy and jet events
at the CERN $\sqrt{s}=14$ TeV $pp$ accelerator, 
the Large Hadron Collider (LHC), will
eliminate the possibility of supersymmetry
accommodating the weak scale naturally. 
The estimated search reaches of gluinos and squarks, which
typically lie above 1 TeV, lie far outside the natural regions
shown in these figures.

\section{Conclusions}
\def\theequation{5.\arabic{equation}}
\setcounter{equation}{0}
 
Searches for supersymmetry have reached a critical 
milestone.  While the absence of observable signals of
supersymmetry at past and present collider experiments has eliminated some
regions of parameters space, experiments have not yet posed any significant
challenge to the viability of the theory.  It is still possible
to find very natural solutions which satisfy all of the quoted, 
experimental bounds on superpartner masses.  
By contrast, as future searches extend beyond the mass ranges
already  explored,
a failure to observe signals of supersymmetry will begin to undermine 
a supersymmetric explanation of why the weak scale remains light.
If LEP-II and the Tevatron's Main Injector era  
do not produce superpartners, 
supersymmetry's explanation of weak scale stability 
will be measurably, but not catastrophically, weakened.  
Conversely, if supersymmetry is relevant to the weak scale,
the chance of observing a signal of superpartner production
at these machines is promising.  If
signals of SUSY are not seen at the CERN Large Hadron Collider LHC 
the stability of the weak scale can no longer be completely explained 
by low energy supersymmetry. 

\section*{Acknowledgments}
This work was supported in part by funds provided by the
U.S. Department of Energy (DOE) under cooperative agreement 
DE-FC02-94ER40818 and by the Texas National Research Laboratory
Commission under grant RGFY932786.

\newpage
\begin{description}
\item[\it Figure 1a:] A comparison of the naturalness contours
$\tilde{\gamma}_{1}$ and $\tilde{\gamma}_{2}$ in terms of the
gluino mass and the mass of the lightest squark belonging to the
first two generations. The dashed and solid curves represent contours
above which the fine-tuning parameters exceed 2.5, 5, 10, and 20
respectively.
\item[\it Figure 1b:] A comparison of the naturalness contours
$\tilde{\gamma}_{1}$ and $\tilde{\gamma}_{2}$ in terms of the
common gaugino mass $m_{1/2}$ and the common scalar mass $m_0$. 
The dashed and solid curves represent contours above which the 
fine-tuning parameters exceed 2.5, 5, 10, and 20 respectively.

\item[\it Figure 2a:] SUSY discovery prospects as a function of the
gluino mass and the lightest squark mass of the first two generations.
Estimates of the physics reach in missing energy and jet events 
(\hbox{/\kern-.6500em $E_{T}$})
at Tevatron runs I and II is compared to naturalness 
contours.  Individual data points represent various 
estimates of the gluino mass reach in the  \hbox{/\kern-.6500em $E_{T}$}
channel for different squark masses.  For comparison,
the dot-dashed curves presents the minimal and maximal
physics reach of chargino production at LEP-II.

\item[\it Figure 2b:] SUSY discovery prospects as a function of the
common gaugino mass $m_{1/2}$ and the common scalar mass $m_0$.
Estimates of the physics reach in missing energy and jet events 
(\hbox{/\kern-.6500em $E_{T}$}) 
at Tevatron runs I and II is compared to naturalness 
contours.  Individual data points represent various 
estimates of the $m_{1/2}$ reach in the 
\hbox{/\kern-.6500em $E_{T}$} channel for different squark masses.
For comparison, dot-dashed (dashed) curves approximate the minimal and maximal
physics reach of chargino (selectron) production at LEP-II.

\item[\it Figure 3a:] The SUSY discovery reach in the tri-lepton
channel (3l) along with naturalness contours in the 
gluino squark mass plane.
Individual data points show estimates of the gluino mass 
reach in the 3l channel for $\mu <0$ and $\tan\beta=2$.

\item[\it Figure 3b:] The SUSY discovery reach in the tri-lepton
channel 3l and naturalness contours in the $m_{1/2}$ $m_0$ plane.
Individual data points show estimates of the gaugino mass 
reach in the 3l channel for $\mu <0$ and $\tan\beta=2$.

\item[\it Figure 4a:] The SUSY discovery reach in the 
\hbox{/\kern-.6500em $E_{T}$} channel for two proposed Tevatron
upgrades compared with naturalness contours in 
the gluino squark mass plane.

\item[\it Figure 4b:] The SUSY discovery reach in the 
\hbox{/\kern-.6500em $E_{T}$} channel for two proposed
Tevatron upgrades compared with naturalness contours in the 
$m_{1/2}$ $m_0$ plane.  

\item[\it Figure 5a:] The SUSY discovery reach in the ($3l$)
channel for two proposed Tevatron upgrades compared with
naturalness contours in the gluino squark mass plane.
Individual data points show estimates of the gluino mass 
reach in the $3l$ channel for $\mu <0$ and $\tan\beta=2$.

\item[\it Figure 5b:] The SUSY discovery reach in the $3l$
channel for two proposed Tevatron upgrades compared with
naturalness contours in the $m_{1/2}$  $m_0$ plane.
Individual data points show estimates of the gaugino mass 
reach in the $3l$ channel for $\mu <0$ and $\tan\beta=2$.

\end{description}

\begin{thebibliography}{99}
%
\bibitem{Witten} E. Witten, Nucl. Phys. {\bf B188}, 513 (1981);
S. Dimopoulos and H. Georgi, Nucl. Phys. {\bf B193}, 150 (1981);
N. Sakai, Z. Phys. {\bf C11}, 153 (1981); R. Kaul, Phys. Lett.
{\bf B109}, 19 (1982).
%
\bibitem{Fine_Tuning} G.W. Anderson and D.J. Casta\~no, 
Phys. Lett. {\bf B347}, 300-308 (1995), hep-ph/9409419; 
Phys. Rev. {\bf D52}, 1693-1700 (1995), hep-ph/9412322.
%
\bibitem{Sensitivity} 
J. Ellis, K. Enqvist, D.V. Nanopoulos, and F. Zwirner,
Mod. Phys. Lett. {\bf A1}, 57 (1986);
R. Barbieri and G. F. Giudice, Nucl. Phys, {\bf B306}, 63 (1988).
%
\bibitem{Wilson} K. Wilson, as quoted by L. Susskind, Phys. Rev. 
{\bf D20}, 2619 (1979); G.'t Hooft, in {\it Recent developments
in gauge theories}, Proceedings of the Cargese Summer Institute,
Cargese, France, 1989, ed by G. 't Hooft et al., NATO Advanced Study
Institute Series B: Physics Vol. 59 (Plenum Press, New
York, 1980)p. 135.  
%
\bibitem{LEPIIa}C. Dionisi {\it et al.}, in ECFA Workshop on LEP 200
(Aachen, West Germany, 1986), edited by A. Bohm and W. Hoogland
(CERN Report No. 87-08, Geneva, Switzerland),Vol. II, p. 380;
H. Baer, M. Chen, C. Dionisi, M. Martinez, and X. Tata, Phys. Rep. {\bf 159},
201 (1988).
%
\bibitem{LEPIIb}J.L. Feng and M. J. Strassler, 
Phys. Rev. {\bf D51}, 4661 (1995); 
N. Oshimoro and Y. Kizukuri, Phys. Lett. {\bf 186B}, 217 (1987).
J.-F. Grivaz, in {\it Proceedings of the 23rd Workshop of the INFN Eloisatron
Project}, {\it eds.} L. Cifarelli and V.A. Khoze (World Scientific, 
Singapore,1993),p. 131.
%
\bibitem{LEPIIc}H. Baer, M. Brhlik, R. Munroe and X. Tata,
Phys. Rev. {\bf D52}, 5031 (1995), hep-ph/9506459.
%
\bibitem{CDFD0}S. Abachi et al., Fermilab PUB-95/057-E, 
submitted to Phys. Rev. Lett,
S. Hagopian FERMILAB-CONF-94-331-E, Oct. 1994.
; F. Abe {\it et. al.} (CDF Collaboration), Phys. Rev. 
Lett. {\bf 69}, 3439 (1992); Phys. Rev. {\bf D45}, 3921 (1992).  
%
\bibitem{BKT}H. Baer, C. Kao, and X. Tata,
Phys. Rev. {\bf D48}, R2978 (1993), 
Phys. Rev. {\bf D48}, 5175  (1993), 
Phys. Rev. {\bf D51}, 2180  (1995). 
%
\bibitem{BCKT}H. Baer, C.-h. Chen, C. Kao, and X. Tata,
Phys. Rev. {\bf D52}, 1565 (1995). 
%
\bibitem{KLMW} T. Kamon, J. Lopez, P. McIntyre and J.T. White,
Phys. Rev. {\bf D50}, 5676 (1994).
%
\bibitem{MKKW} S. Mrenna, G.L. Kane, G.D. Kribs, and J.D. Wells,
hep-ph/9505245.
%
\bibitem{DPF}H. Baer et. al. hep-ph-9503479
To be published in the DPF Working Groups' Reports ( World Scientific,
Singapore);
T. Barklow, S. Dawson, H. Haber, and J. Siegrist, 
{\it ibid}, hep-ph/9505296;
M. Drees and S. Martin hep-ph  Report of the DPF Working Group.
{\it ibid. }, hep-ph/9504324, and references therein.
%
\bibitem{GF} C. Albajar et al, UA1 Collab., Phys. Lett. 
{\bf B198}, 261 (1987); G. Farrar,
Phys. Rev. {\bf D51}, 3904 (1995).
%
\end{thebibliography}
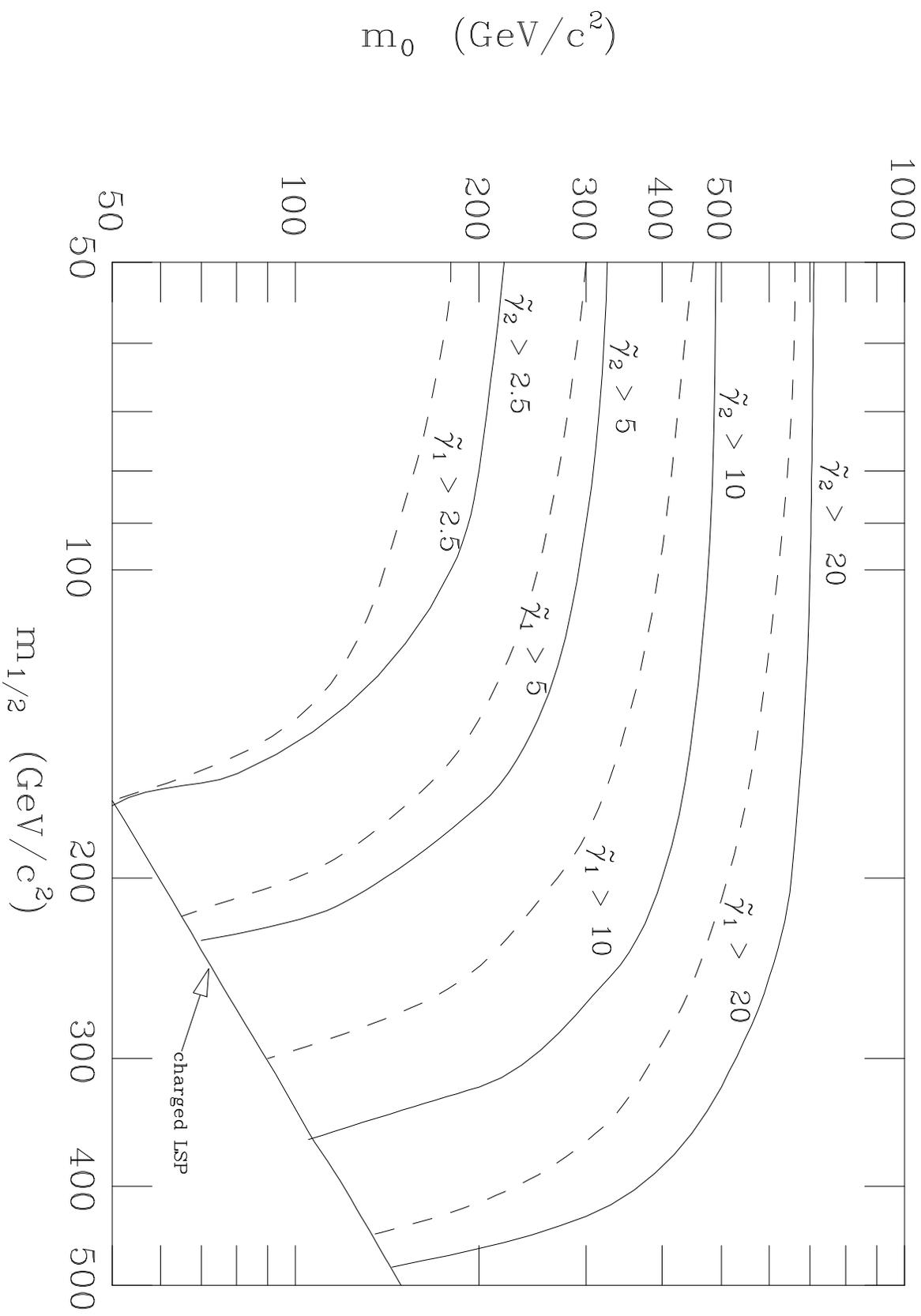
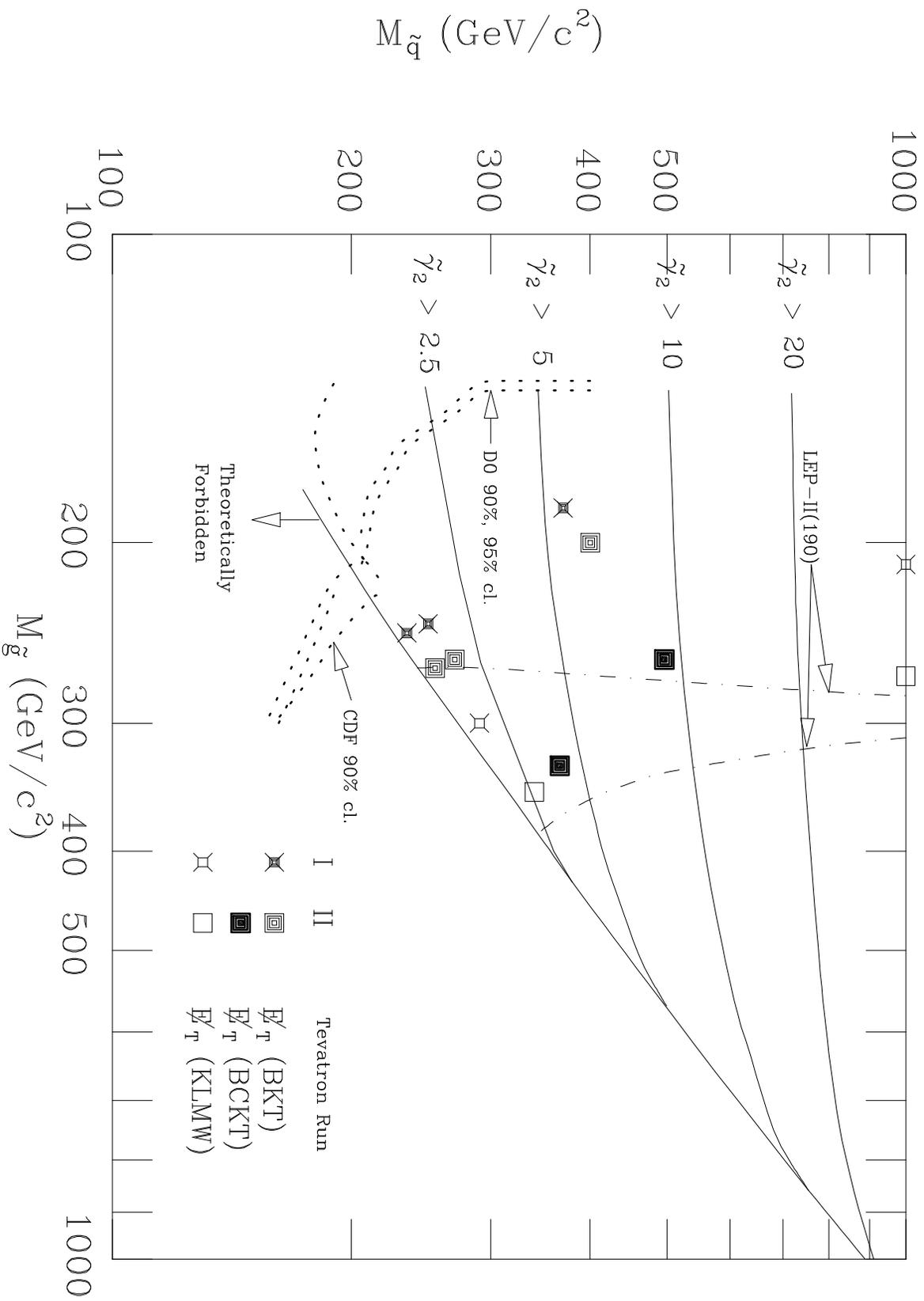
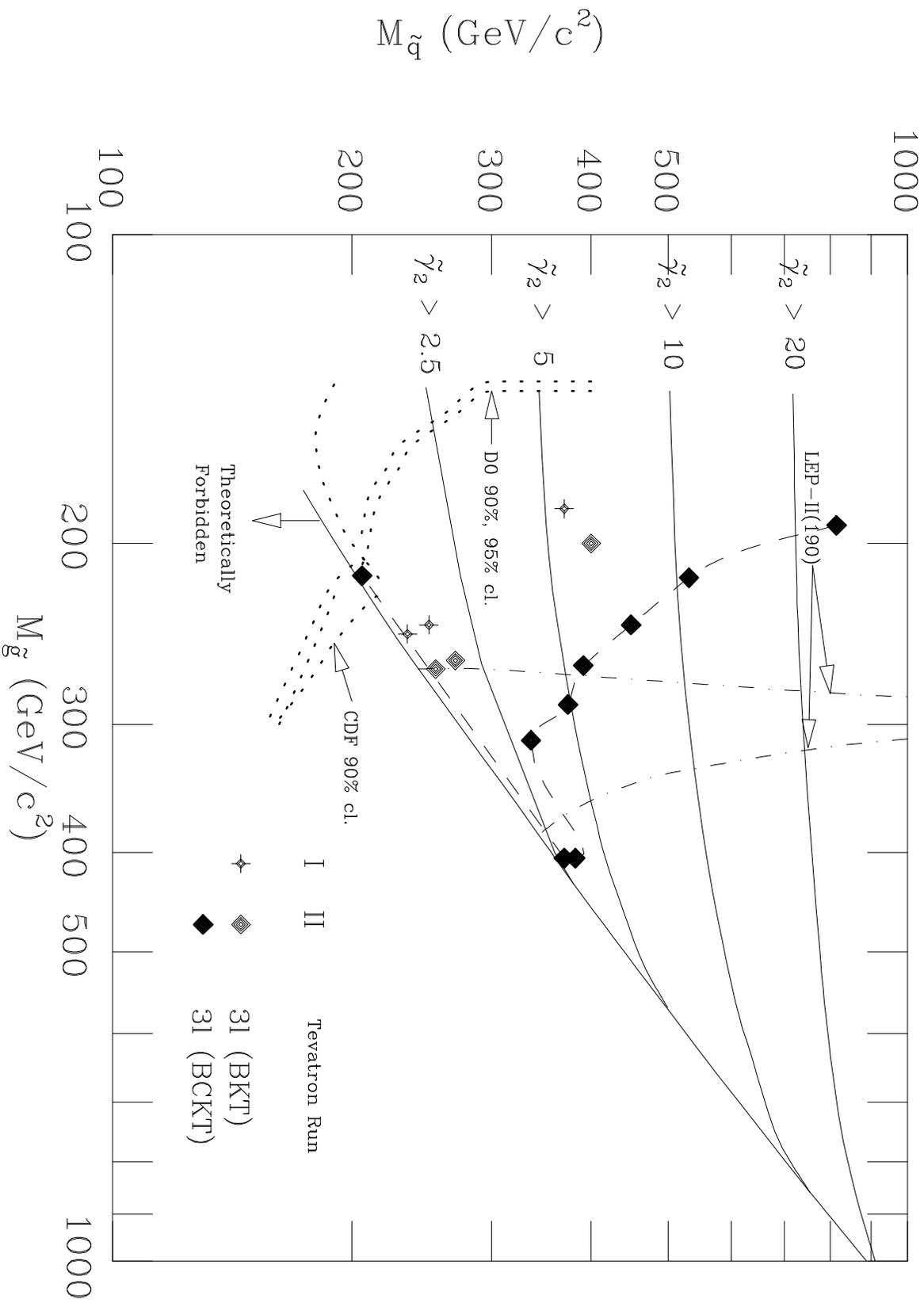
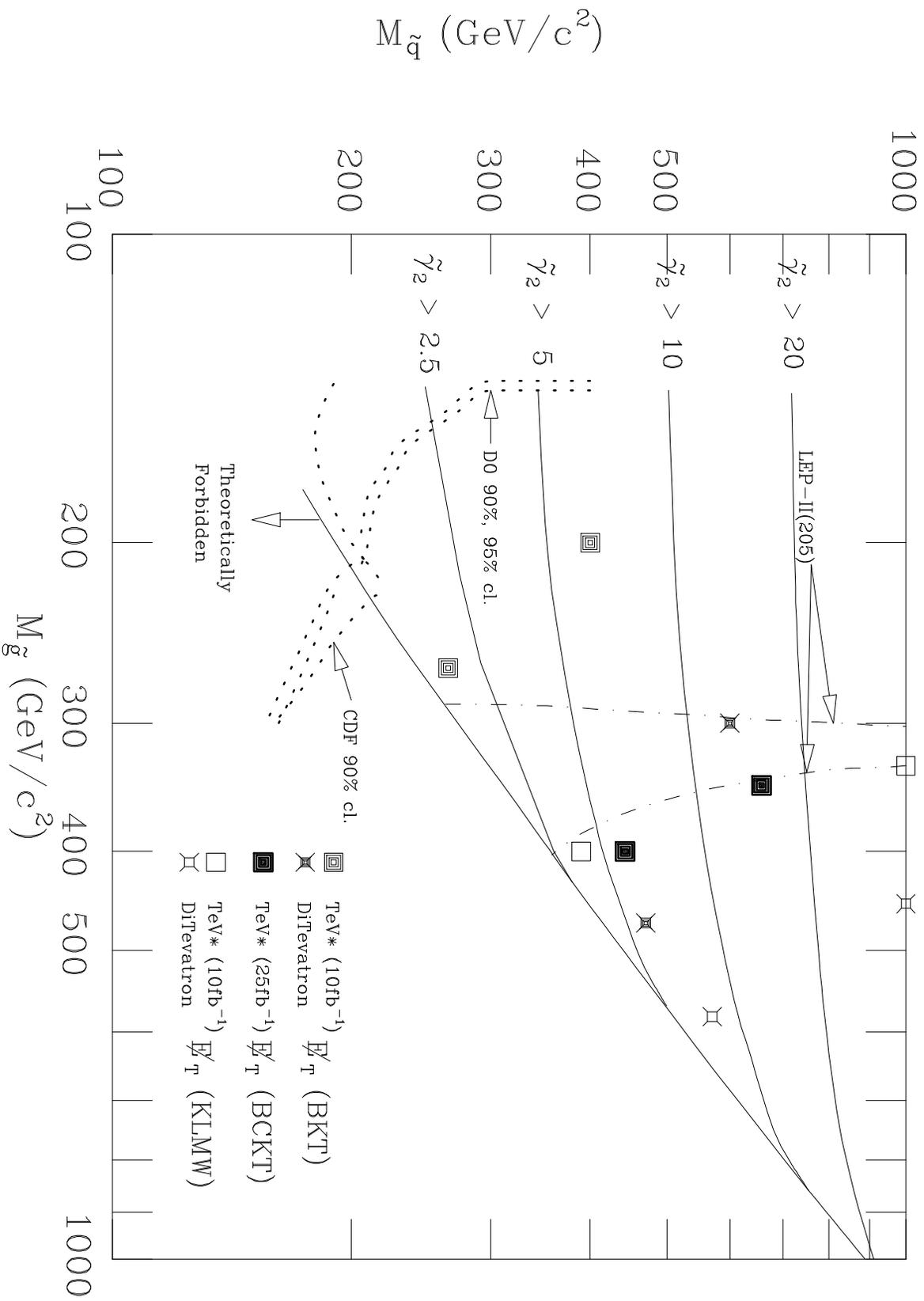
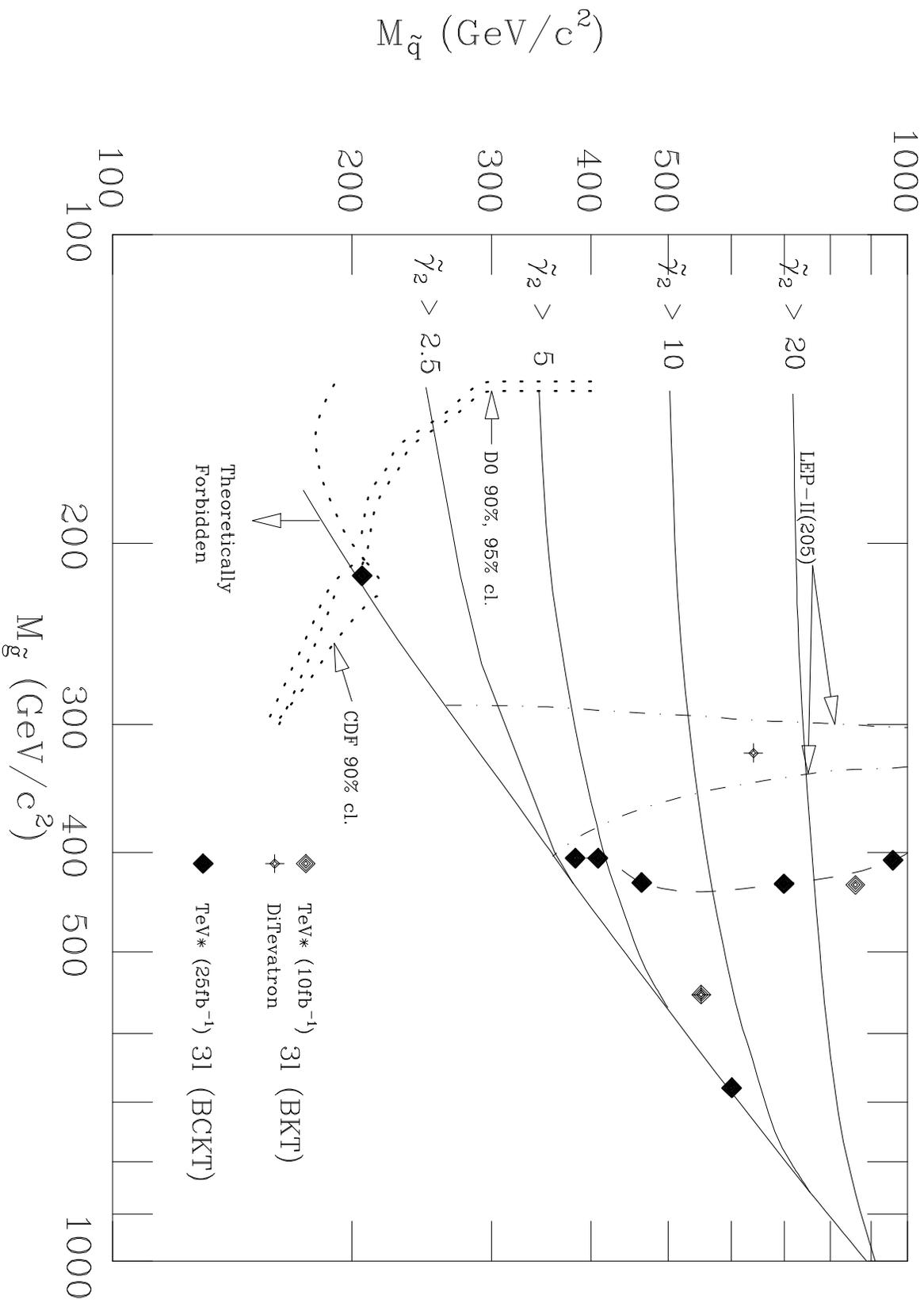
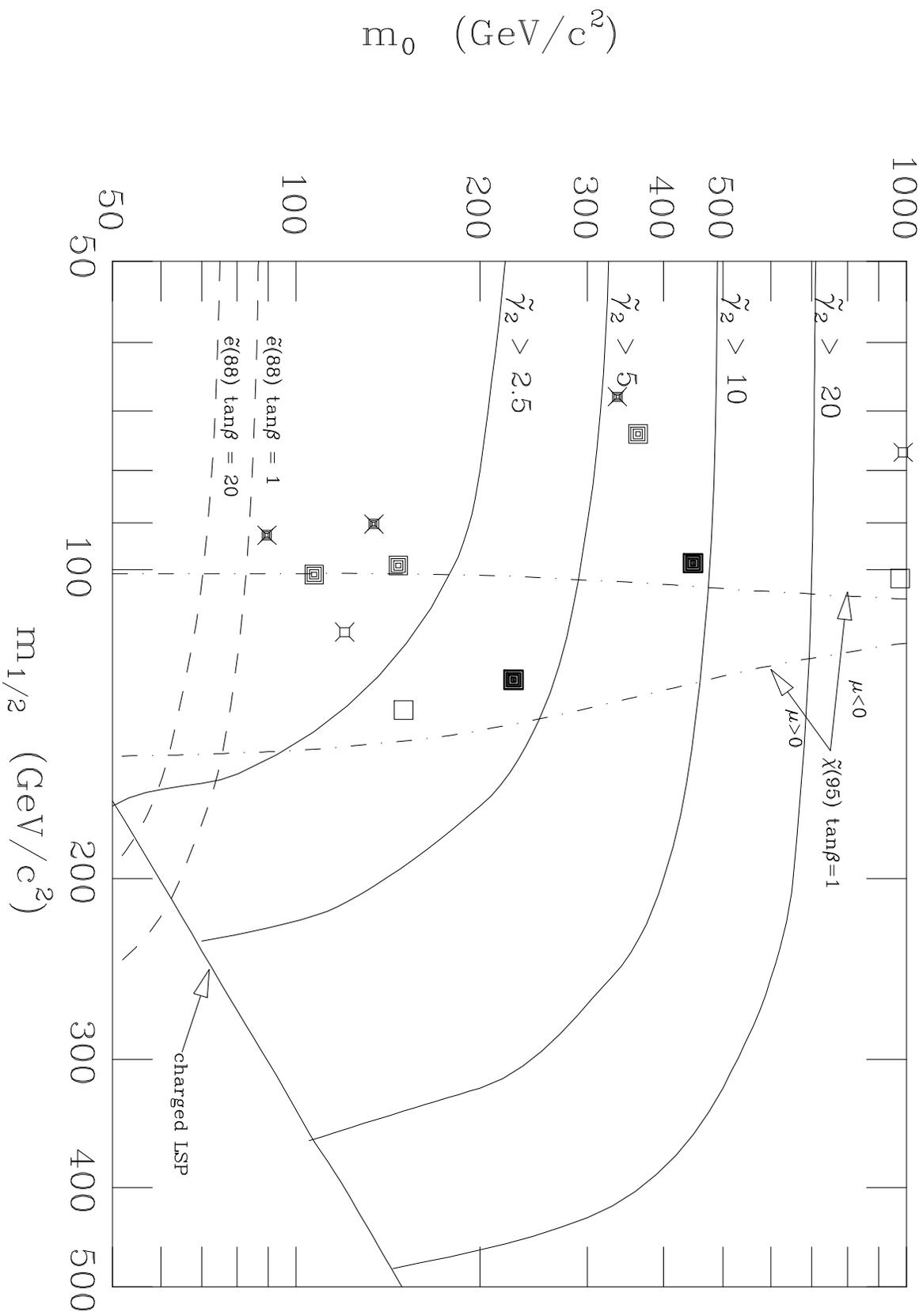
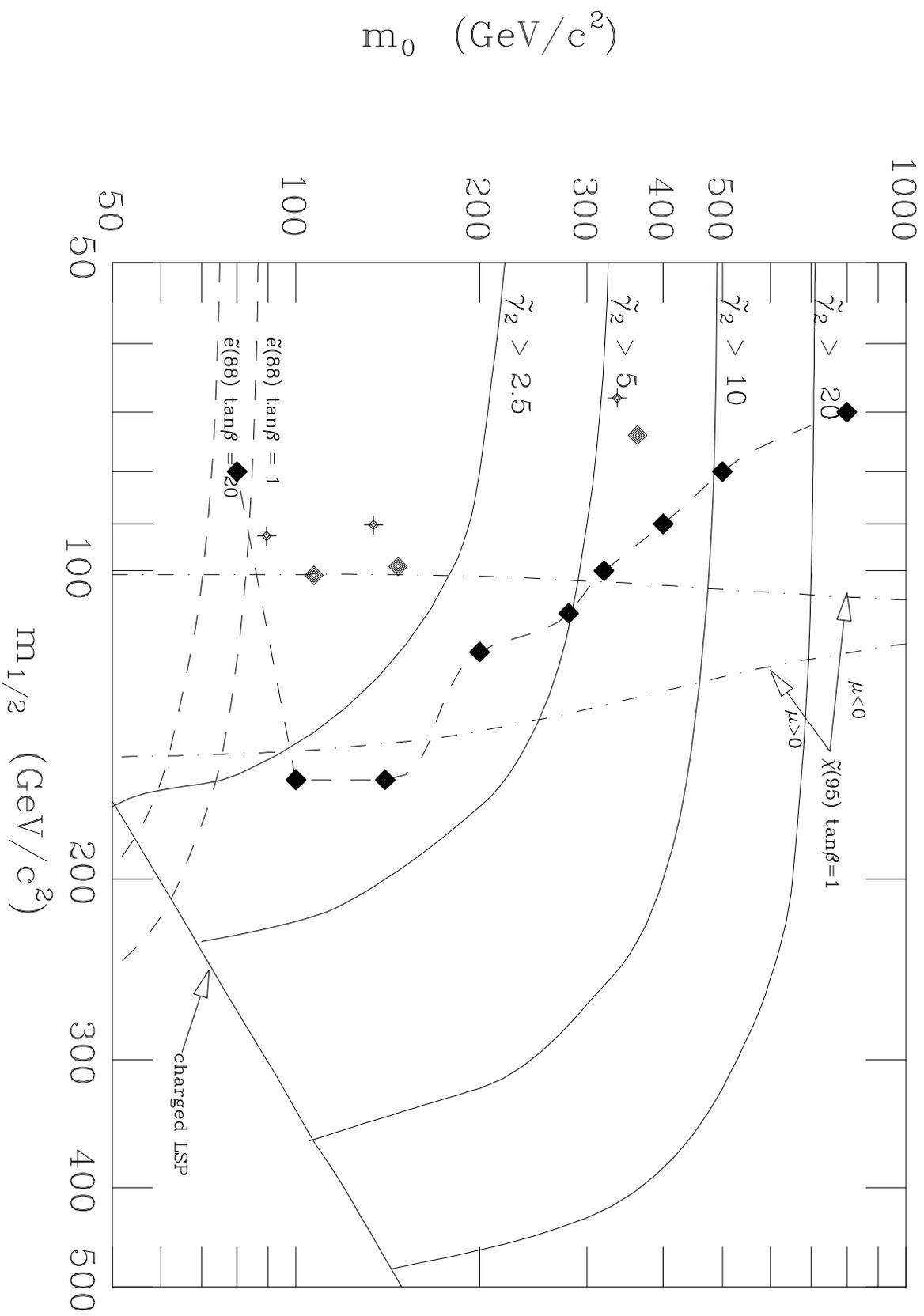
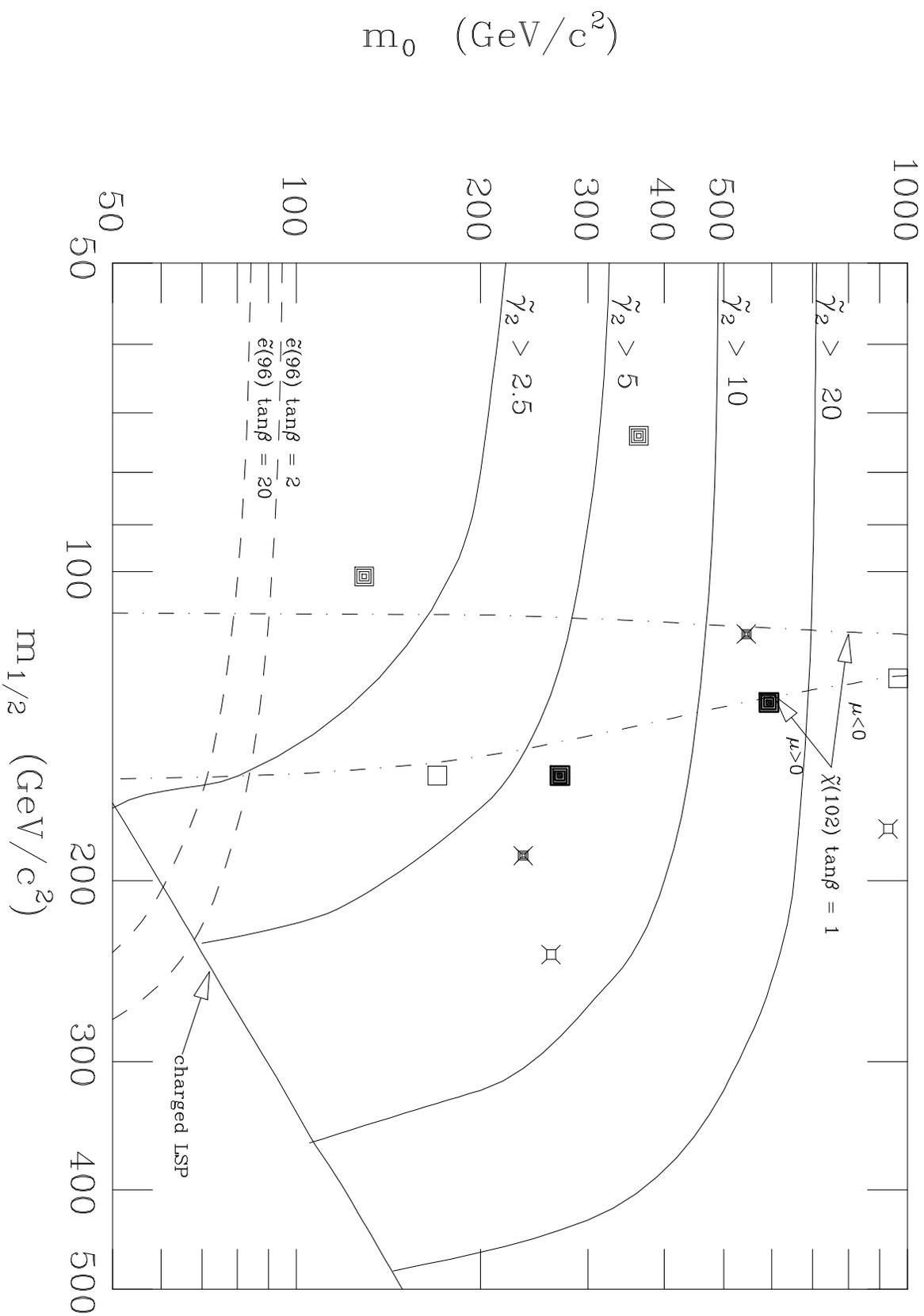
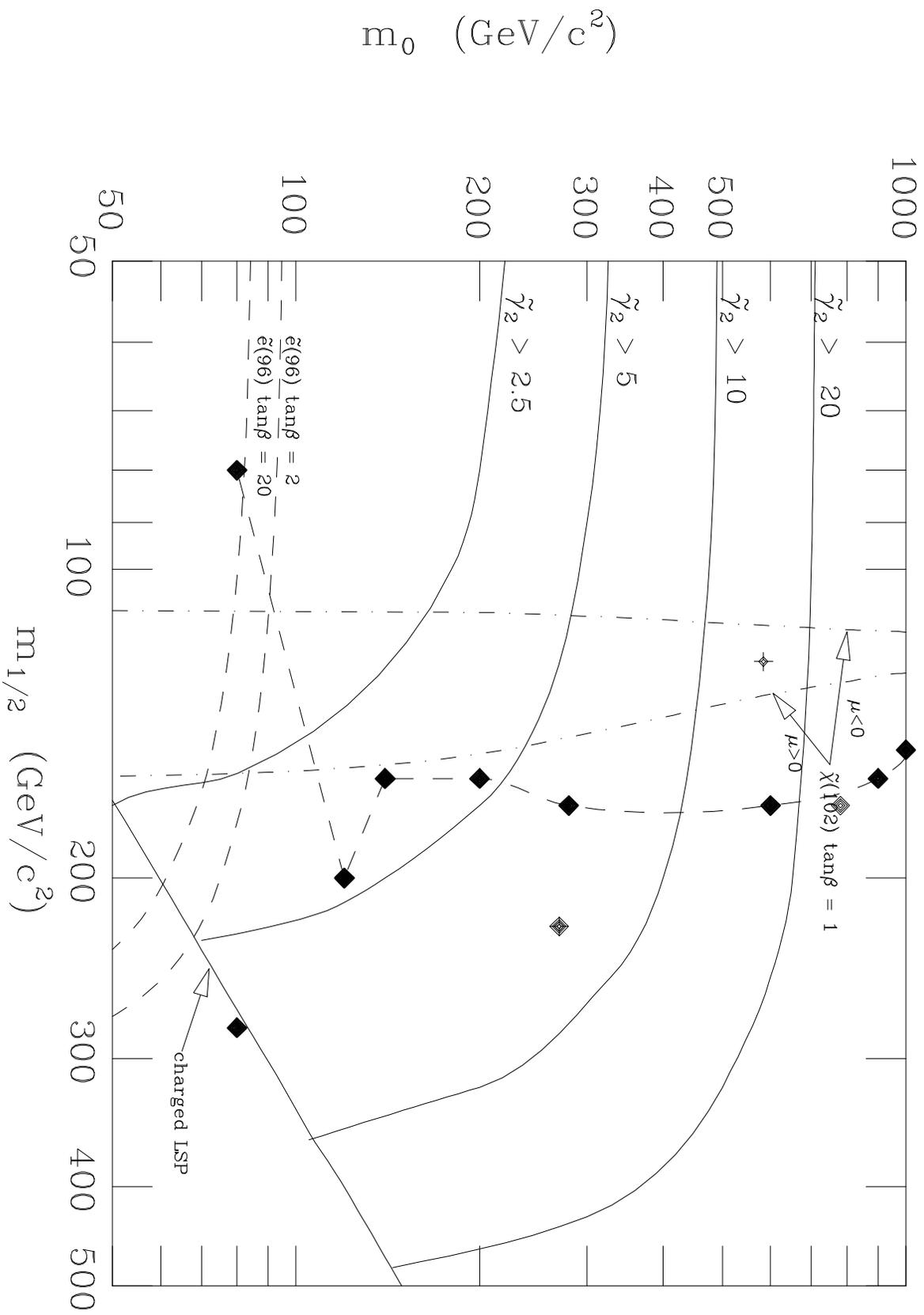
\end{document}